\documentclass[11pt]{article}
\usepackage{latexsym}
\usepackage{epsfig}
\begin{document}
\newcommand{\bea}{\begin{eqnarray}}
\newcommand{\eea}{\end{eqnarray}}
\newcommand{\dif}{{\rm d}}
\newcommand{\ct}{\cos \theta}
\newcommand{\st}{\sin \theta}
\newcommand{\sd}{\sin^2 \theta}
\newcommand{\scu}{\sin^3 \theta}
\newcommand{\cd}{\cos^2 \theta}
\newcommand{\ccu}{\cos^3 \theta}
\newcommand{\cq}{\cos^4 \theta}
\newcommand{\cc}{\cos^5 \theta}
\newcommand{\csx}{\cos^6 \theta}
\newcommand{\csv}{\cos^7 \theta}
\newcommand{\tth}{\tan \theta}
\newcommand{\ctt}{\cot \theta}
\newcommand{\ECM}{\em Departament d'Estructura i Constituents de la
Mat\`eria
                  \\ Facultat de F\'\i sica, Universitat de Barcelona \\
                     Diagonal 647, E-08028 Barcelona, Spain\\
                                     and           \\
                                    I. F. A. E.  \\
                                     and \\
                     Syracuse University  \\
		     Physics Department\\
		     201 Physics Building \\
		     Syracuse 13244-1130 NY, USA
}
\def\thefootnote{\fnsymbol{footnote}}
\pagestyle{empty}
{\hfill \parbox{6cm}{\begin{center} UB-ECM-PF 99/05\\
                                    February 1999
                     \end{center}}}
\vspace{1.5cm}

\begin{center}
\large{{\bf $\eta$ and $\eta'$ hadronic decays in $U_L(3) \otimes U_R(3)$ 
Chiral Perturbation Theory.}}
\end{center}
\vskip .6truein
\centerline {P. Herrera-Sikl\'ody
\footnote {herrera@ecm.ub.es}}
\vspace{.3cm}
\begin{center}
\ECM
\end{center}
\vspace{1.5cm}

\centerline{\bf Abstract}
\bigskip
The decays $\eta / \eta' \rightarrow \pi \pi \pi$ and
$\eta' \rightarrow \eta \pi \pi$
are studied up to leading
and next-to-leading order within the framework
of $U_L(3) \otimes U_R(3)$ Chiral Perturbation Theory.
The analysis incorporates important features of the $\eta-\eta'$ 
system, such as 
the contribution of the glueball $\alpha G\tilde G$ 
due to the axial anomaly and
$\eta_0 / \eta_8$ mixing.
One-loop corrections, which are third-order 
contributions according to the combined chiral and $1/N_c$ expansion,
are not included.
Reasonably good results are obtained in most cases.

\bigskip

\bigskip


\newpage
\pagestyle{plain}

\section{Introduction}

Chiral perturbation theory has proved to be a good tool to describe
the dynamics of the low-energy region of QCD, where the natural
degrees of freedom are the eight Goldstone bosons (pions, kaons and 
$\eta$) associated with the spontaneous breaking of  
$SU_R(3) \otimes SU_L(3)$ chiral symmetry. If a large number
of colors $N_c$ is allowed, the theory can be enlarged to 
the so-called $U_R(3) \otimes U_L(3)$ Chiral Perturbation Theory, 
that includes
a ninth particle --- the $\eta'$.

The $\eta \rightarrow \pi \pi \pi$ decays have drawn
a lot of attention from theoreticians since they provide a measure
of the isospin violation due to the quark masses: the main contribution
to the amplitude is proportional to $m_d-m_u$. In principle, this
process can be analyzed in the simpler octet theory. 
However, the $\eta$ particle
is actually a superposition of the $SU(3)$ singlet $\eta_8$ and
the $U(1)$ singlet $\eta_0$, and mixing is seemingly important 
($\theta \approx -20^{\circ}$), so a model including this effect
is expected to give better results. 

The measured values \cite{pdb} for $\eta$ decays are the following:
\bea
\Gamma (\eta\rightarrow \pi_0 \pi_0 \pi_0) &=&  379 \pm 40 \; {\rm eV} \ , \nonumber \\
\Gamma(\eta\rightarrow \pi_0 \pi_+ \pi_-) &=& 274 \pm 33 \; {\rm eV} \ ,\nonumber \\
\nonumber \\
r=\frac{\Gamma^{exp} (\eta\rightarrow \pi_0 \pi_0 \pi_0)}
{\Gamma^{exp} (\eta\rightarrow \pi_0 \pi_+ \pi_-)} &=& 1.35 \pm 0.05 . \nonumber
\eea

\vspace{1em}

The $U(3)$ theory can be also used to study 
$\eta'$ decays. Two different channels 
will be analyzed in this paper: $\eta' \rightarrow \pi\pi\pi$ and
$\eta' \rightarrow \eta\pi\pi$.

The $\eta' \rightarrow \pi\pi\pi$ 
transition is also an isospin violating process.
It is however {\em not} one of the dominant decay channels for
the $\eta'$, as happened in the corresponding $\eta$ decay. 
As a consequence, the branching ratios
associated to these decays are more difficult to determine 
experimentally
because they stem from a small fraction
of the total observed events, so the uncertainties are unfortunately  
higher:
\bea
\Gamma^{exp} (\eta'\rightarrow \pi_0 \pi_0 \pi_0) &=&  311 \pm 77 \; {\rm eV}   \ .
\nonumber \\
\Gamma^{exp} (\eta'\rightarrow \pi_0 \pi_+ \pi_-) &<&  1005 \; {\rm eV} \ ,
\nonumber \\
r &>& 0.3 \nonumber
\eea   

The $\eta' \rightarrow \eta\pi\pi$ transition is in contrast one of the
most important decay channels for the $\eta'$.    
The measured rates are:
\bea
\Gamma^{exp} (\eta'\rightarrow \eta \pi_0 \pi_0) &=&  42.0 \pm 6.0 \; {\rm keV}  \ , \nonumber \\
\Gamma^{exp} (\eta'\rightarrow \eta \pi_+ \pi_-) &=&  88.9 \pm 10.0 \; {\rm keV}  \ ,\nonumber \\
r = \frac{\Gamma (\eta'\rightarrow \eta \pi_0 \pi_0)}
{\Gamma (\eta'\rightarrow \eta \pi_+ \pi_-)}&=& 2.1 \pm 0.5\ .\nonumber
\eea   

\vspace{2em}

In general, the description of these decays involves the estimation of the
following amplitudes:
\bea
\langle \, {\cal P}_1 {\cal P}_2 {\cal P}_3 \, | \, {\cal P}_0 \, 
\rangle = i \, 
(2 \, \pi)^4 \,  
\delta^4(p_f-p_i)\, A_{{\cal P}_0 \rightarrow
{\cal P}_1 {\cal P}_2 {\cal P}_3} 
(s,t,u)\; , \nonumber 
\eea
where
\bea
s=(p_0-p_1)^2 \quad ; \quad t=(p_0-p_2)^2 \quad ; \quad
u=(p_0-p_3)^2  \quad \mbox{($p^{\mu}_i$ is the momentum of particle ${\cal P}
_i$)}\ .
\nonumber
\eea
The decay rates in the center-of-mass reference frame 
require a phase space integral of the squared amplitudes over
some region ${\cal R}$, defined by the kinematic restrictions of a three body 
decay.

Some general issues can be inferred from symmetry considerations.
In the isospin limit (and in the absence of electromagnetic interactions)
the $\eta/\eta' \rightarrow \pi\pi\pi$ decays 
are strictly forbidden by Bose symmetry. Therefore, they
must originate in isospin-violating terms,
which are proportional to $m_u-m_d$.
One can see \cite{gl2} that the pions must emerge in an
$I=1$ configuration. This $\Delta I=1$ selection rule can be used to prove 
a relation between the two decay channels:  
\bea
A_{000}(s,t,u)\, =\, A_{0+-}(s,t,u)\, +\, A_{0+-}(t,u,s)\, +
\, A_{0+-}(u,s,t) \ .
\label{amplitudes}
\eea

If one assumes that the dominant decay channel is the vectorial one
(by vector-meson dominance \cite{vmd}) 
where $\eta/\eta' \rightarrow \pi \rho \rightarrow \pi \pi \pi $,
one can assume the amplitudes to be flat, 
moment independent functions:
$A_{0+-}(s,t,u) \approx A_{0+-}(s) \approx A_{0+-}(M^2_{\rho}) 
\equiv {\cal A}$.
Under this assumption,
the ratio between the charged and the neutral channels can be easily
estimated:
\bea
r=\frac{\Gamma (\eta\rightarrow \pi_0 \pi_0 \pi_0)}
{\Gamma (\eta\rightarrow \pi_0 \pi_+ \pi_-)} \approx
\frac{\frac{1}{3!}\int_{\cal R} |3 {\cal A}|^ 2} {\int_{\cal R} |{\cal A}|^ 2}
=  1.5 \ .
\label{ratio} 
\eea
Notice that phase space corrections due to $M_{\pi^0}\neq M_{\pi^+}$ 
have been neglected (the region of integration ${\cal R}$ is the same in
the numerator and the denominator). 
They can easily be included within the present 
approximation, but turn out to increase rather than decrease the ratio 
---the main corrections to it must come from the inclusion of other decay
channels.

The isospin symmetry constraints on the $\eta' \rightarrow \eta\pi\pi$
system produce more restrictive results. In this case, 
the wave function must be symmetric under the exchange of pions
only. Besides, 
the total isospin for the three-particle state must be equal to zero.
When this is taken into account, the expansion in terms of 
Clebsch-Gordan coefficients leads to a simple relation:
\bea
A_{\eta' \rightarrow \eta \pi_+\pi_-}(s,t,u) = 
A_{\eta' \rightarrow \eta \pi_0\pi_0}(s,t,u) \ . \nonumber
\eea

The amplitudes being equal, the rates will be identical except
for the combinatorial 1/2! pre\-factor in the neutral case, 
so the ratio must be equal to 2. Phase-space corrections produce a
deviation from this value.

\vspace{1em}

A first approach to the $\eta$ decays can be done in 
the usual $SU(3)$ Chiral Perturbation Theory. The leading-order results
are definitely too small. The ${\cal O}(p^ 4)$ result is
closer to the experimental data, because the unitary
corrections due to the final-state interactions 
are surprisingly large, especially in the I=0 and 1, S-wave $\pi\pi$ channels. 
These results (collected in section 5) 
seem to indicate that the $\eta \rightarrow \pi \pi \pi$
decays are dominated by the (well-known) vectorial resonance and
by an intermediate low-mass scalar resonance, 
the celebrated $\sigma$ particle \cite{harada, oller}. 

Nevertheless, as mentioned before, 
there is another possible reason why the $SU(3)$ Chiral 
Perturbation Theory prediction
fails: the physical $\eta$ particle is not one of the
states in the octet, but a superposition 
of $\eta_8$  and $\eta_0$. The $\eta_0$, on the other
hand, is a mixture of the pseudoscalar quark current 
and the gluonic state $G \tilde{G}$ due to
the axial anomaly. All this should be taken into account in order to get
a more careful description of the $\eta$ meson \cite{picheta}.

\vspace{2em}
\section{$U_L(3) \otimes U_R(3)$ Chiral Perturbation Theory}
\vspace{1em}

$U_L(3) \otimes U_R(3)$ Chiral Perturbation Theory is an appropriate 
tool to deal with the $\eta -\eta' -\pi$ transitions. 
In the chiral limit $m_q=0$, the QCD
Lagrangian is invariant under the flavor group $SU_L(3)\otimes 
SU_R(3)$. As a consequence, the vector and axial 
currents are classically conserved. 
However, the symmetry observed in nature 
is $SU_V(3)$, and only approximatively.
This means that the classical largest symmetry must somehow be 
spontaneously broken: $SU_L(3)\otimes 
SU_R(3) \rightarrow SU_V(3)$. The remaining 
$SU_V(3)$ symmetry is also slightly broken
due to the different masses of the different quarks:
\bea
\partial_{\mu} V_{\alpha}^{\mu} &=& 0\;+\;{\cal O}(m_{q}-m_{q'})\; , 
\nonumber \\
\partial_{\mu} A_{\alpha}^{\mu} &=& 0\;+\;{\cal O}(m_q)\ , \hspace{10em}  
\alpha \neq 0 \ . \nonumber
\eea
The low-energy spectrum of QCD is made of the 
well-known octet of Goldstone bosons corresponding to the eight broken axial 
symmetries.

The singlet part of the flavor group deserves a separate discussion,
because even in the chiral limit, the axial current is not conserved
due to the presence of the axial anomaly. 
Nevertheless, the anomalous terms
are proportional to the inverse of the number of colors $N_c$, so  
$U_A(1)$ can be indeed considered as a good symmetry 
if one also allows $N_c$ to be large:
\bea
\partial_{\mu} A_0^{\mu} &=& 0\;+\;{\cal O}(m_q)\;
+\;{\cal O}(\frac{1}{N_c})\; .\nonumber
\eea 
In this case, one can also think of this symmetry as being spontaneously
broken and a ninth Goldstone boson appears.

Following the spirit of Chiral Perturbation Theory, one is led to build
an effective theory containing nine pseudoscalar particles, associated
with the spontaneous 
breaking of the symmetry $U_L(3)\otimes U_R(3) \rightarrow U_V(3)$ and
conveniently collected for this purpose in a 3 $\times$ 3 matrix $U$:
\begin{eqnarray}
U(x) = \exp \left( i \sum_{\alpha =0}^{8} 
\frac{\lambda_{\alpha} \phi_{\alpha}(x)}{f} \right), 
\qquad\mbox{where}\quad \{ \lambda _{\alpha} \}_{\alpha =0, ..., 8}
\quad\mbox{are the $U(3)$ generators.}\quad 
\nonumber
\end{eqnarray}
As usual the external sources $s_{\alpha}(x)$, $p_{\alpha}(x)$, 
$v_{\alpha}^{\mu}(x)$ and $a_{\alpha}^{\mu}(x)$,
coupled to the vector, axial, scalar and
pseudoscalar QCD currents respectively, are introduced 
in order to generate Green's functions for the QCD currents and because their
behavior with respect to the flavor group transformations, 
taken from QCD, guarantees the reproduction of
the QCD symmetry breaking pattern in the effective theory.
For instance the explicit symmetry breaking effects due to
$m_q \neq 0$ are introduced by freezing the value of 
the scalar source $s(x)=2\,B\,{\cal M}$,
where $B$ is a constant and
${\cal M}=diag(m_u,\, m_d,\,m_s)$ is the quark-mass matrix. 
Similarly, in the $U(3)\otimes U(3)$ case, 
the source $\theta(x)$, coupled to 
$:\alpha_s G(x)\tilde{G}(x):$ in the QCD Lagrangian, provides an
excellent tool to keep track of the effects of the anomaly.

In terms of these sources, the divergence of the singlet
axial currents reads:
\bea
\frac{\delta \, {\cal S}_{QCD}}{\delta \, 
\partial_{\mu} a_0^{\mu}(x)} &=& 2\, \sqrt{\frac{2}{n_f}}\, \Big(\,  
\, \sum_{\alpha} \ {\cal M}_{\alpha} \frac{\delta \, {\cal S}_{QCD}}{\delta \, 
p_{\alpha}(x)}\; - \; 
n_f \, \frac{\delta {\cal S}_{QCD}}{\delta \, \theta(x)}\, \Big)\, \; .
\label{anomaly}
\eea
The non-conservation of $A_0^{\mu}$
originates from two different sources: firstly from the quark masses
through the pseudoscalar currents coupled to $p_\alpha (x)$ and secondly 
from the anomaly through the current coupled to $\theta(x)$.
When $S_{QCD}$ is replaced by
$S_{\chi {\rm PT}}$, (\ref{anomaly}) is automatically satisfied (the 
sources have been defined to do so!). Recall however that, in this case, 
the equation refers to infinite series
of the fields, because the different QCD currents are 
represented by infinite
series of pseudoscalar mesons. 

These two building blocks (the matrix $U$ and external sources) and the 
symmetry constraints allow for an infinite number of operators in the
Lagrangian, but they can be classified depending on the
associated power of energy (one derivative $\sim p \sim m_q$ ).
The complete ${\cal O}(p^0)$, ${\cal O}(p^2)$ and ${\cal O}(p^4)$
Lagrangians were given in \cite{nosaltres}. The problem is that
each term in the Lagrangians can be multiplied by an (almost) arbitrary
function of a special combination of the singlet
field and the external source $\theta$ given by 
$X= i\ \frac{\sqrt{6}}{f}\ \phi_0 + i\ \theta$. Fortunately, the 
$N_c$ counting allows one to
think of these functions as infinite series in $X$ whose coefficients
would be suppressed in one power of $1/N_c$ for each power of $X$.
Due to discrete symmetry restrictions, these series must be either
made of exclusively even or odd powers of $X$.
Following the notation introduced in \cite{nosaltres}, $W_k(X)$ (k=0 to 6)
will refer to the arbitrary 
functions that multiply ${\cal O}(p^0)$ and ${\cal O}(p^2)$ operators, and
$L_k(X)$ (k=0 to 57) will be used for the ones that multiply ${\cal O}(p^4)$ 
operators. The coefficients for the expansion in powers of $X$ are
the parameters that are used in practice:
\bea
W_k(X) &=& \frac {f^2}{4}
\left( v_{k,0} + v_{k,2} X^2 + v_{k,4} X^4 + ...\right), \quad k \neq 3\; ,
\nonumber \\ 
W_3(X)&=&-i \frac {f^2}{4}
\left( v_{3,1} X + v_{3,3} X^3 + \; ...\right),\nonumber \\ 
L_k(X) &=& L_{k,0} + L_{k,2} X^2 + L_{k,4} X^4 + ..., \quad \mbox{for
even operators},\nonumber \\ 
L_k(X) &=& L_{k,1} X + L_{k,3} X^3 + ..., \quad \qquad \mbox{for
odd operators}.
\nonumber
\eea

This strengthens the idea that the only possible way of 
working within this theory is thus through
the use of two simultaneous expansions in powers of masses and 
momenta and in powers
of $1/N_c$. Luckily 
simple arguments based on the
value of the $\eta'$ mass \cite{meta3} 
suggest that both expansions can be merged 
by assuming: 
$p^2 \; \sim M^2 \; \sim \; m_q \; \sim \; 1/N_c \; \sim \; \delta$ 
($M$ being the typical meson mass). 

\vspace{1em}

Under this assumption, the leading-order Lagrangian reduces to three terms:
\bea
{\cal L}_{LO} &=& \frac{f^2}{4}\Big( -\ v_{0,2} X^2 + 
\langle D_{\mu}U^{\dagger}D^{\mu} U \rangle + 
\langle U^{\dagger} \chi+ \chi^{\dagger} U \rangle \Big)\; ,
\nonumber
\eea
where:
\bea
D_{\mu}\ U&=& \partial_{\mu}\ U - i\ r_{\mu}\  U +i\  U\  l_{\mu}\; ,
\nonumber \\
D_{\mu}\ U^{\dagger}&=& \partial_{\mu}\ U^{\dagger} -
i\ l_{\mu}\  U^{\dagger} +i\  U^{\dagger}\  r_{\mu}\; ,\nonumber \\
\chi&=& 2\ B\ (s+i\ p)\; ,\nonumber \\
r_{\mu} &=& v_{\mu}+a_{\mu} \ , \nonumber \\
l_{\mu}&=& v_{\mu}-a_{\mu} \ ,
\nonumber
\eea
and
\bea
v_{0,2}\sim \frac{1}{N_c} \; , \qquad f \sim \sqrt{N_c} \; , \qquad B\sim N_c^0.
\nonumber
\eea

Notice that the actual order in $\delta$ of any particular computation
will depend on the number of external fields, because each field
carries a $1/f \sim N_c^{-1/2} \sim \delta^{1/2}$ factor. 

A simple dimensional analysis shows that 
the one-loop diagrams will always be suppressed by a factor $M^2/f^2$ ---which
is ${\cal O}(\delta^2)$ according to our choice of $\delta$. 
As a consequence, the next-to-leading order Lagrangian, that is
suppressed with respect to the leading one by a factor $\delta$ only, can
also be treated classically because quantum corrections would only appear in
the third order in the expansion.

\vspace{1em}
The next-to-leading Lagrangian involves new ${\cal O}(p^2)$ and ${\cal O}(p^4)$
terms:
\bea
{\cal L}_{NLO}&=& {\cal L}_{LO} \nonumber  \\ 
&+& \frac{f^2}{4} \
\Big(\ v_{3,1}\ X \langle U^{\dagger} \chi- \chi^{\dagger} U \rangle + 
v_{4,0} \langle U^{\dagger}\ D_{\mu}U \rangle 
\langle U^{\dagger}\ D^{\mu}U \rangle \nonumber  \\ 
&+& i\ v_{5,0} \langle U^{\dagger}\ D_{\mu}U \rangle D^{\mu}\theta -
v_{6,0} D_{\mu}\theta  D^{\mu}\theta \Big)    +
 \sum_{j} \ L_j \ O_j  \; ,
\label{lag}
\eea
where $j$ labels the ${\cal O}(p^4)$ operators $O_j$ whose 
coupling constants $L_j$
are of ${\cal O}(N_c)$ (the number of required ${\cal O}(p^4)$ operators
depends on a choice that will be discussed in section 4). 
Either $v_{4,0}$, $v_{5,0}$ or $v_{6,0}$ can be set to zero
by means of a change of variables; we shall use $v_{5,0} = 0$
and $ v_{3,1},\ v_{4,0},\ v_{6,0} \sim \delta$.

\vspace{2em}
\section{Masses and interpolating field in the isospin violating case}
\vspace{2em}

The first step in the calculation is necessarily the identification of
the physical states contained in the theory, i.e., the diagonalization
of the  ${\cal O}(\phi ^2)$ effective action. This was already done
in the isospin limit in \cite{meta3}. However, since
the $\eta/\eta' \rightarrow \pi \pi \pi$ decays are isospin-violating 
processes, the amplitudes for these processes must be evaluated in the case
$m_u\neq m_d\neq m_s$. 

In the next-to-leading order, the two-point functions are described
by an ${\cal O}(\phi ^2)$ and ${\cal O}(\delta^2)$ effective action
given by (\ref{lag}):
\bea
{\cal S}_{{\delta} ^2} =
\frac {1}{2}\int  \; \dif ^4 x \left( \;
\partial_\mu \phi_a\, {\cal A}_{ab} \,
\partial^\mu \phi_b - \phi_a {\cal B}_{ab} \, \phi_b \right),
\label{action}
\eea
In the isospin-violating case,
\bea
{\cal A}=I +\Delta A +\Delta d A\; , \hspace {2em}
{\cal B}=2\, B\, m\, (D+d D +\Delta D+\Delta d D)\;  ,  \hspace {2em}
\nonumber
\eea
where $m=(m_u+m_d)/2$. The suffix $\Delta$ indicates the next-to-leading
terms. The terms with the suffix $d$ are proportional to $m_d-m_u$.
The matrices $D$, $\Delta A$ and $\Delta D$ 
\cite{meta3} do not include isospin-breaking contributions
\footnote {Notice that we have chosen a slightly different set of parameters 
than in \cite{meta3}:
now $v_{5,0}=0$ and $v_{4,0}\neq 0$.}:

\begin{eqnarray}
D_{11}=D_{22}=D_{33}=1\qquad ,\qquad
D_{44}=D_{55}=D_{66}=D_{77}=1+{x\over 2}\ ,\nonumber\\
D_{88}=1+{2\over 3}\ x\quad ,\quad
D_{08}=-{{\sqrt 2}\over 3}\ x\quad ,\quad 
D_{00}=1+{1\over 3}\ x-{3\over 2}\ {v_{02}\over mB} \ ,
\nonumber
\end{eqnarray}
The strange quark mass is introduced through the quantity $x=(m_s-m)/m$.

The ${\cal O}(p^4)$ operators involved in the 2-point functions are $O_5$ and
$O_8$. There are also two ${\cal O}(p^2)$ contributions, because $v_{3,1}$
and $v_{4,0}$ are of ${\cal O}(1/N_c)$. All these terms contribute to
the ${\cal O}(\delta^2)$ corrections:

\begin{eqnarray}
&&\Delta  A_{11}=\Delta A_{22}=\Delta A_{33}=\frac{16\ L_{5,0}}{f^2}\ B\, m
 \ ,\nonumber\\
&&\Delta A_{44}=\Delta A_{55}=\Delta A_{66}=\Delta A_{77}=
\frac{16 \ L_{5,0}}{ f^2}\  B\, m\left(1+{x\over 2}\right)  \ , \nonumber \\
&&\Delta  A_{88}= \frac{16\ L_{5,0}}{ f^2}\  
B\, m\left(1+{2\over 3}x\right)
 \ ,\nonumber \\
&&\Delta A_{08}=-  \frac{16\ L_{5,0}}{f^2}\ \frac{\sqrt 2}{3}\ B\, m\ x
 \ ,\nonumber \\
&&\Delta A_{00}=\frac{16\ L_{5,0}}{ f^2}\  B\, m
\left(1+{1\over 3} x\right) \  -3\ v_{4,0}\ ,
\nonumber
\end{eqnarray}

and:

\begin{eqnarray}
&&\Delta D_{11}=\Delta D_{22}=\Delta D_{33}=\frac{32 \ L_{8,0}}{f^2}\ B\, m
 \ ,\nonumber\\
&&\Delta D_{44}=\Delta D_{55}=\Delta D_{66}=\Delta D_{77}=
\frac{32 \ L_{8,0}}{f^2}\ B\, m\left(1+x+{1\over 4}\ x^2\right) \ ,\nonumber\\
&&\Delta  D_{88}=\frac{32 \ L_{8,0}}{f^2}\ B\, m\left(1+{4\over 3}\ x
+{2\over 3} \ x^2\right)
 \ ,\nonumber\\
&&\Delta  D_{08}=-  \frac{32 \ L_{8,0}}{ f^2}\ {{\sqrt 2} \over 3}\  B\, m
\ x\ (2+x)+{\sqrt 2}\  v_{3,1}\  x
 \ ,\nonumber\\
&&\Delta  D_{00}=\frac{32 \ L_{8,0}}{f^2} \ B\, m 
\left(1+{2\over 3}\  x +{1\over 3}\  x^2\right)-
2\  v_{3,1} \ (3+x) \ .
\nonumber
\end{eqnarray}
\vspace{1em}

When $m_u\neq m_d$, the following (leading and next-to-leading) 
pieces must also be considered:
\bea
&&\Delta dA_{44} = \Delta dA_{55} \; =\; - \Delta dA_{66}\; =\; 
-\Delta dA_{77} \; =\;   
\frac{8\, L_{5,0}}{ f^ 2}\ B\, (m_u-m_d)
\; , \nonumber \\
&&\Delta dA_{38} = \frac{8\, L_{5,0}}{ f^ 2}\,  \frac{1}{\sqrt{3}}\, 
B\, (m_u-m_d) 
\; , \nonumber \\
&&\Delta dA_{03} = \frac{8\,  L_{5,0}}{ f^ 2} \, \sqrt{\frac{2}{3}}\, 
B\, (m_u-m_d)
\; , \nonumber 
\eea
\bea
&&dD_{44} =dD_{55}\; =\; -dD_{66}\; =\; - dD_{77}\; =\;
\frac{m_u-m_d}{4\ m}
\; , \nonumber \\
&&dD_{38}= \frac{1}{\sqrt{3}}\, \frac{m_u-m_d}{2\ m}
\; , \nonumber \\
&&dD_{03} = \sqrt{\frac{2}{3}}\, \frac{m_u-m_d}{2 \ m}
\; , \nonumber 
\eea

\bea
&&\Delta dD_{44} = \Delta dD_{55}\; =\; -\Delta dD_{66}\; =\; 
-\Delta dD_{77}\; =\;
\frac{L_{8,0}}{ f^ 2} \,  B\, \frac{(m_u-m_d)^2}{m^2}
\; , \nonumber \\
&&\Delta dD_{38} = \frac{32\, L_{8,0}}{ f^ 2} \, \frac{1}
{\sqrt{3}}\, B\,  (m_u-m_d) \; , \nonumber \\
&&\Delta dD_{03}=\frac{ 32\, L_{8,0}}{ f^ 2}\,  \sqrt{\frac{2}{3}}\,  B\, 
(m_u-m_d)\; .
\nonumber
\eea

A simple change of variables:
\bea
\psi=(I\, +\, \frac{1}{2}\,  \Delta A\,  +\, \frac{1}{2} \, 
\Delta d A )\, \phi\;  
\label{psi}
\eea
provides the correct prefactor in
the kinetic term 
and reduces the calculation of the masses to an eigenvalue problem. 
The matrix to be diagonalized is: 
\bea
2\, B\, m\, \Big( D+\Delta D - \frac{1}{2} \{ \Delta A,D \} + d D + \Delta dD-
\frac{1}{2} \{ \Delta d A,D \}- \frac{1}{2} \{ \Delta A,d D \}\Big)\; .
\eea
This matrix is diagonal in the kaon and the charged-pion sector, so
the interpolating fields and the
particle masses can be written straightforwardly as:
\bea
\pi_+ = -\frac{1}{\sqrt 2} (\psi_1+ i\psi_2) \; , &&
\pi_- = \frac{1}{\sqrt 2} (\psi_1- i\psi_2)\; , \nonumber \\
K_+ = -\frac{1}{\sqrt 2} (\psi_4+ i\psi_5) \; , &&
K_- = \frac{1}{\sqrt 2} (\psi_4- i\psi_5) \; , \nonumber \\
K_0 = -\frac{1}{\sqrt 2} ( \psi_6+ i\psi_7) \; , &&
\bar{K}_0 = \frac{1}{\sqrt 2} (\psi_6- i\psi_7) \; . \nonumber
\eea
\bea
&&M^2_{\pi_+} = M^2_{\pi_-} = B\,(m_u+m_d) \Big( 1+8 \, B\, (m_u+m_d)
\frac{2\, L_{8,0}-L_{5,0}}{f^2} \Big) \; , \nonumber \\
&&M^2_{K_+} = M^2_{K_-} = B\, (m_u+m_s) \Big( 1+8 \, B\,   (m_u+m_s) 
\frac{2\, L_{8,0}-L_{5,0}}{f^2} \Big) \; , \nonumber \\
&&M^2_{K_0} = M^2_{\bar{K_0}} = 
B\, (m_d+m_s) \Big( 1+8 \, B\,   (m_d+m_s) 
\frac{2\, L_{8,0}-L_{5,0}}{f^2} \Big) \; . \nonumber
\eea

$\pi_0$, $\eta$ and $\eta'$ are related to $\phi_3$,
$\phi_0$ and $\phi_8$ by a rotation $S$. In the $m_u=m_d$ case, the rotation
mixes $\eta$ and $\eta'$ only:
\bea
S=\pmatrix{1&0&0\cr 0&\ct&\st\cr
0&-\st & \ct\cr}\; .
\nonumber
\eea
The isospin-violating terms will be considered as
small corrections to be treated perturbatively. To first order in
this expansion, the relation between the original fields and the
eigenvectors is given by a matrix $F^d$
\footnote{
These expressions are to be understood in the sense of perturbation
theory:
\bea
F^d \,=\, S^{-1}(I+\frac{1}{2} \Delta A +
\frac{1}{2} \Delta d A ) + dT\, S^{-1}\, (I+\frac{1}{2} \Delta A)  +
\Delta dT\, S^{-1} + {\cal O}(\delta^2, d^2)\, . \nonumber
\eea}:
\bea
\varphi^d_P \, &=& (F^d)_{P\alpha} \; \phi_{\alpha} \; ,\qquad 
\mbox{with} \quad
F^d= (S^d)^{-1}(I+\frac{1}{2} \Delta A +
\frac{1}{2} \Delta d A ) \,
, \nonumber \\
\mbox{and} \quad
S^d &=& S\, (I-dT- \Delta dT) \; .
\nonumber
\eea 
The matrices $dT$ and $dDT$ are defined by:
\bea
dT_{PQ} &=& \frac{2\, B\, m} {M_P^2- M_Q^2}\,\langle \, \varphi_P \, | \; d D
|\, \varphi_Q \, \rangle,  \qquad \qquad  (\varphi= S \psi) \ ,
\nonumber \\
\Delta dT_{PQ} &=& \frac{2\, B\, m} {M_P^2- M_Q^2}\, 
 \langle \, \varphi_P \, | \; \Delta dD
 - \frac{1}{2} \{ \Delta d A,D \}- \frac{1}{2} \{ \Delta A,d D \}  \, 
|\, \varphi_Q \, \rangle \, . \nonumber
\eea
The only non-vanishing elements in these matrices are:
\bea
dT_{\pi \eta} &=& -dT_{\eta \pi} \; = \; \epsilon_0 
(\ct - \sqrt{2}\st) \, , \nonumber \\
dT_{\pi \eta'} &=& -dT_{\eta' \pi} \; = \; 
\frac{\Delta}{M_{\eta'}^2-M^2_{\pi}}\,  \epsilon_0 \,  
(\st + \sqrt{2}\ct)
\, , \nonumber \\
\Delta dT_{\pi \eta} &=& -\Delta dT_{\eta \pi} \; = \; 
(\epsilon_{3,8} \ct - \epsilon_{0,3} \st)
 , \nonumber \\
\Delta dT_{\pi \eta'} &=& -\Delta dT_{\eta' \pi}\; = \; 
\frac{\Delta}{M_{\eta'}^2-M^2_{\pi}}
\; (\epsilon_{3,8} \st + \epsilon_{0,3} \ct), 
\label{dT}
\eea
where
\bea
\epsilon_{3,8} &=&  \epsilon_0  \, 16 \, (M_{\pi}^2-M_K^2) \, 
\frac{2 \, L_{8,0}-L_{5,0}}{f^2}  \; , \nonumber \\
\epsilon_{0,3} &=&  \epsilon_0 \, \sqrt {2} \, \Big( - 3\, v_{3,1} +
12\, v_{0,2}\, \frac{L_{5,0}}{f^2}+16\, (M_{\pi}^2-M_K^2) \,
\frac{2 \, L_{8,0}-L_{5,0}}{f^2} + \frac{3}{2}\, v_{4,0}\Big) \; , \nonumber \\
\Delta &=& M_{\eta}^2-M_{\pi}^2 \; . \nonumber 
\eea

The dimensionless quantity $\epsilon_0$ is a good measure of 
the isospin-breaking perturbation since
\bea
B(m_d-m_u)=-\epsilon_0 \, \Delta
 \, \sqrt{3}\, \Big(1-16 \, M_K^2 \, 
\frac{2 \, L_{8,0}-L_{5,0}}{f^2} \Big) \, .
\nonumber
\eea
and can be estimated in terms of the observable quantities $\Delta$ and
$M_1^2$ (\ref{m1}):
\bea
\epsilon_0= \frac{M_1^2}
{\sqrt{3} \, \Delta } \; , \qquad
M_1^2=(M_{K_0}^2 -M_{K_+}^2) -(M_{\pi_0}^2-M_{\pi_+}^2) \; .
\label{m1}
\eea
The combination of masses $M_1^2$ isolates the QCD  
isospin-breaking effect due to the quark masses, 
because the electromagnetic contribution to the mass
splitting is the same for the kaons and the pions and will cancel out.

It can be checked that the leading-order contributions to the
matrix $F^d$ do indeed match the
diagonalization given in \cite{leut2}. 

As a consequence of (\ref{dT}) being the only non-vanishing terms 
the first isospin-violating corrections to 
$M^2_{\pi^0}$, $M^2_{\eta}$ and $M^2_{\eta'}$ are of 
${\cal O}(\epsilon_0^2)$ which goes beyond our working precision 
and will be therefore neglected.

In any case, for the isospin-violating decays that are the subject of
this paper, the amplitude is proportional to $\epsilon_0$ so 
only the complete isospin-violating eigenstates and the complete
quark mass matrix are required for the
calculation. The free parameters in 
the theory can be estimated
in the isospin limit since any $\epsilon_0$ correction would give
a second-order contribution of ${\cal O}(\epsilon_0^2)$ to the amplitude.

\vspace{2em}
\section{Four-point processes in $U_L(3) \otimes U_R(3) \ \chi {\rm PT}$}
\vspace{2em}
To leading order, the only two relevant four-field terms in the Lagrangian
are formally the same as those that appear in the $SU(3)$ theory:
\bea
{\cal L}_{\phi^4}&=& -\frac{1}{6\, f^2}\, f_{abr}\, f_{cdr}\, 
\phi_a \, \partial_{\mu}\phi_b \,\phi_c \, \partial_{\mu}\phi_d \,+
\, \frac{1}{24\, f^2} d_{abr}\, d_{res}\, d_{scd} \, \phi_a
\, \phi_b \, \phi_c \, \phi_d \,2\, B\,  {\cal M}_e \; . \nonumber
\eea

\vspace{2em}
The next-to-leading ${\cal O}(\delta^2)$ contributions can originate from:

\begin{enumerate}
\item[a)] terms with constants of ${\cal O}(1/N_c)$
from the ${\cal O}(p^0)$ Lagrangian;
\item[b)]  terms with 
constants of ${\cal O}(N_c^0)$
from the ${\cal O}(p^2)$ Lagrangian; 
\item[c)]   terms with 
constants of ${\cal O}(N_c)$ from the
${\cal O}(p^4)$ Lagrangian.
\end{enumerate}

The $N_c$-power counting for the coupling constants in the model was
studied in \cite{nosaltres}. According to this work
there can be no ${\cal O}(p^0)$ correction, 
and there will be only one more ${\cal O}(p^2)$ contribution, associated with
the coupling constant $v_{31}$. The next chiral order looks
less encouraging: at first sight 
one might think that nine independent ${\cal O}(p^4)$ terms have to
be included ($O_1$, $O_2$, $O_3$, $O_5$, $O_8$, $O_{13}$, $O_{14}$, $O_{15}$ 
and $O_{16}$). This problem can be dodged by noticing that the constants $L_i$ 
associated to most of them are ${\cal O}(N_c)$ due to the
contribution of the term $O_0$ that was eliminated through the
Cayley-Hamilton theorem. A more convenient set of
independent operators can be chosen by eliminating $O_{16}$ instead. 
In this case, the only operators left in the list are $O_0$, $O_3$, $O_5$
and $O_8$, 
because all the other terms are suppressed by a factor $1/N_c$ or more.

The new set of coupling constants $\{ M_i\} $ is related to the old one 
$\{ L_i\} $ . In 
particular, for the first three of them, one obtains :
\bea
M_{1,0}=L_{1,0} - \frac{M_{0,0}}{2} \quad  ; \quad M_{2,0}=L_{2,0} - M_{0,0}
\quad  ; \quad
M_{3,0}=L_{3,0}+2 M_{0,0}\; . \nonumber
\eea
One can use the first two relations to get an estimate of the
new constant $M_{0,0}$. 
$M_{1,0}$ and $M_{2,0}$ are expected to be both of ${\cal O}(N_c^0)$,
and thus negligible in front of $M_{0,0}$, $L_{1,0}$ and $L_{2,0}$.
Furthermore, as shown in \cite{int}, $L_{1,0}$, $L_{2,0}$ and $L_{3,0}$ are
equal to the corresponding $L_{1}$, $L_{2}$ and $L_{3}$ from the
octet theory up to one-loop corrections ---which is beyond our
working precision. $L_{3,0}$ will be therefore directly borrowed from the
$SU(3)$ model
(see, for instance, \cite{pichlh} and references therein):
\bea
L_{3,0} &=& (-3.5 \pm 1.1)\ \cdot 10^{-3} \; .
\label{l30}
\eea
$L_{1}$ and $L_{2}$ can be used to fix the new constant:
\bea
M_{0,0} = \frac{2}{3} (L_{1,0}+L_{2,0}) + {\cal O}(N_c^0) \simeq 
(1.2 \, \pm\,  0.4)\; 10^{-3} \; .
\nonumber
\eea
Another way to estimate $M_{0,0}$ is given by the 
QCD bosonization models \cite{boson}:
\bea
M_{0,0} = \frac{N_c}{192 \pi^2} \approx 1.58 \cdot 10^{-3} \; .
\nonumber
\eea
This value seems more reliable since bosonization models
have proved to give excellent results for the ${\cal O}(p^4)$ constants
(except for the operators that contain explicit symmetry breaking because
then the results are model-dependent, but this is not the case for $O_0$).

Yet a third possible source to 
fix $M_{0,0}$ can be used: 
it turns out that the most important
contributions to the decay $\eta' \rightarrow \eta \pi \pi$ come from $O_0$, 
so the experimental data for this process can be used to determine $M_{0,0}$ 
(see section 7 for details). The quoted error is the one 
induced by the experimental uncertainties:
\bea
M_{0,0}&=& (1.54 \pm 0.1)\ \cdot 10^{-3} \; .
\label{m00}
\eea 
This value is in good
agreement with the theoretical estimations discussed above. 
It is worth pointing out that $M_{3,0}$ turns out to
be quite small. Using the central value in (\ref{m00}), we take:
\bea
M_{3,0} = -0.4 \cdot 10^{-3} \; . 
\label{m30}
\eea

\vspace{1em}

The other constants that appear in the calculation are $B \, m$,
$x$
\footnote{Recall that, to first order in
$m_u-m_d$, the isospin-breaking effect is an overall factor in the
amplitude.}
, $f$, $L_{5,0}$, $L_{8,0}$, $v_{3,1}$, $v_{0,2}$ and $v_{4,0}$. Their 
values are fixed by the masses and decay constants $M_{\pi}$,
$M_{K}$, $M_{\eta}$, $M_{\eta'}$, $f_{\pi}$,
$f_{K}$, $f_{\eta}$ and  $f_{\eta'}$. Within the required precision and 
in terms of the mixing angle $\theta$, the first correction can be written
in terms of measurable quantities by using the following identities: 
\bea
&&\Delta_M={8\over f^2} (M^2_K-M^2_\pi) (2 L_{8,0}- L_{5,0}) \ =\ 
\frac{M^2_\pi + 3 M^2_\eta -4 M^2_K +3(M^2_{\eta'}-M^2_\eta)\ 
\sin^2 \theta}{4\ (M^2_K-M^2_\pi)} \ , \nonumber \\
&&\Delta_N= 3\ v_{3,1} - {12\over f^2}\ v_{0,2}\ L_{5,0} \ =\ 
1+\frac{3}{4 \sqrt{2}}\ \frac{(M^2_{\eta'}-M^2_\eta)}
{(M^2_K-M^2_\pi)}\ \sin 2\theta 
\ , \nonumber \\
&&\frac{L_{5,0}}{f^2}= \frac{1}{4\ (M^2_K-M^2_\pi)}
\Big(\frac{f_K}{f_{\pi}}-1\Big)\ , \nonumber \\
&&v_{4,0}=-\frac{2}{3}\Big( \frac{f_{\eta}+f_{\eta'}}{f_{\pi}}-2
\Big) \ . 
\label{cons1}
\eea
The parameters from the leading-order Lagrangian must be
evaluated up to ${\cal O}(\delta)$ corrections:
\bea
B\ m &=& M^2_{\pi}\ (1-\frac{M^2_{\pi}}{M^2_K-M^2_\pi}\ \Delta_M)
\ , \nonumber \\
x &=& 2 \ \frac{M^2_K}{M^2_\pi}(1- \Delta_M)-2\ (1+ \Delta_M)\ , \nonumber \\
f &=& f_{\pi} \Big( 1-\frac{4\ L_{5,0}}{f^2}\  M^2_{\pi}\Big) \ , \nonumber \\
-3\ v_{02} &=& \Big(M^2_{\eta'} -\frac{2 M^2_K+M^2_\pi}{3} \Big) (1-3\ v_{40})
+ \frac{2 \sqrt{2}}{3}\  (M^2_K-M^2_\pi)\  ( 1+\Delta_M-\Delta_N 
-\frac{3}{2} \ v_{4,0})\ \tth  \nonumber \\ &-&
\frac{2}{3} \Big((M^2_K-M^2_\pi)\ \Delta_M +
(2 M^2_K+M^2_\pi)\ \Delta_N \Big) \ .
\label{cons2}
\eea

As shown in \cite{meta3}, the fitting does not fix the value of the 
mixing angle $\theta$. One observes, however, that the corrections on
$M^2_{\eta}$ and $M^2_{\eta'}$ given by $\Delta_M $ and $\Delta_N$ 
are minimized for $\theta$ between $-20^{\circ}$ and $-22^{\circ}$. 
Furthermore, this 
{\em minimum sensitivity} prediction agrees with the
experimental data and is reasonably close to the 
leading-order prediction ($\theta \approx -21.7^{\circ}$), as
expected if the $U(3)$ expansion is to make sense. 


\vspace{2em}
\section{The $\eta \rightarrow \pi \pi \pi$ decays}
\vspace{2em}

A first estimation of the decay rates 
can be given by the current algebra \cite{wein} or by
the leading order ${\cal O}(p^2)$ in $SU(3)$ Chiral 
Perturbation Theory \cite{gl}, \cite{gl2}. According to
this theory, the amplitude is:
\bea
A_{0+-}(s,t,u)= -\frac{\epsilon_0}{f_{\pi}}\, (s-\frac{4}{3} M_{\pi}^2)\, .
\nonumber
\eea
Electromagnetic contributions to the decay amplitude need not
be included to leading and next-to-leading order
in the low-energy expansion, as 
argued in \cite{gl}. The only relevant QED effect in this case is the
splitting between the masses of the charged particles, which is taken into
account through the definition of $\epsilon_0$ (\ref{m1}). 
\vspace{1em}

The numerical results are too small:
\bea
\Gamma^{SU(3)} (\eta\rightarrow \pi_0 \pi_0 \pi_0) &=& 100 \; {\rm eV} \ ,\nonumber \\
\Gamma^{SU(3)} (\eta\rightarrow \pi_0 \pi_+ \pi_-) &=& 66 \; {\rm eV} \ ,\nonumber \\
r&=&1.51\; . \nonumber 
\eea
The predicted rates are off by a factor of four. The branching ratio r is 
however not that bad. This is not such an amazing feature, since it has been
shown that the ratio is related to the isospin selection rules
(see comment on section 1 and appendix A). In any case,
it seems to indicate that whatever is missing 
in both calculations is essentially a multiplicative correction that
cancels out in the ratio.
\vspace{2em}

The next-to-leading ${\cal O}(p^4)$ calculation in $SU(3)$ Chiral 
Perturbation Theory was carefully
analyzed and presented in \cite{gl2}:
\bea
\Gamma^{SU(3)} (\eta\rightarrow \pi_0 \pi_0 \pi_0) &=&  229 \; {\rm eV} \ ,\nonumber \\
\Gamma^{SU(3)} (\eta\rightarrow \pi_0 \pi_+ \pi_-) &=& 160 \; {\rm eV} \ ,\nonumber \\
r&=&1.43 \; . \nonumber 
\eea
The new terms come from the
${\cal O}(p^4)$ Lagrangian and from one-loop diagrams. An unexpectedly large
contribution arises from the unitarity correction term, especially from
the pieces that correspond to $\pi\pi$ final-state interaction
\footnote {The importance of the so-called {\em rescattering} effects
had been already pointed out in some previous work made in the
context of current algebra which attempted to improve the initial 
result by including the $\pi\pi$ final-state 
interactions through the imposition of unitarity and analyticity 
\cite{rt}.}. Technically, these contributions turn out to be
so important because
the one-loop diagram integrals are plagued with infrared divergences
that produce a significant enhancement of the quark-mass perturbation. 

\vspace{1em}

\subsection{Leading order in $U_L(3) \otimes U_R(3) \ \chi{\rm PT}$}
\vspace{1em}

As shown in section 3, the leading-order Lagrangian required to
describe this process is very similar to the Lagrangian used in $SU(3) \ 
\chi {\rm PT}$. The difference with respect to that case lies in the 
mixing effects. The anomaly appears only 
indirectly through
the contribution of $v_{02}$ to $M_{\eta}^2$ and $\theta$.
\bea
A_{0+-}&=& -\frac{\epsilon_0}{f_{\pi}}\, \Big( \,  (\ct-\sqrt{2} \st)\,  
(s-\frac{4}{3}M^2_{\pi})  + \frac{M^2_{\pi}}{3}\, \tth \,
(\st+4\, \sqrt{2}\, \ct)\Big)\ .
\label{eopm} 
\eea
The last term is proportional to $M^2_{\pi}$ so it produces a 
small correction to the final result.
The mixing effects reduce essentially to a factor of $\ct -
\sqrt{2} \st$ (=1.45 for $\theta = -21.7^{\circ}$) \cite{leut2,fg}.

The predicted rates are higher than in the octet model:
\bea
\Gamma (\eta\rightarrow \pi_0 \pi_0 \pi_0) &=& 180  \; {\rm eV} ,\nonumber \\
\Gamma (\eta\rightarrow \pi_0 \pi_+ \pi_-) &=& 121 \; {\rm eV} ,\nonumber \\
r&=&1.49\; . \nonumber 
\eea

\vspace{1em}
\subsection{Next-to-leading order in $U_L(3) \otimes U_R(3) \ \chi {\rm PT}$ }
\vspace{1em}

The final result for the squared amplitude (consistently expanded into 
leading-order piece + corrections) 
is a very lengthy expression that can be found in the
appendix.
In particular, most of the free parameters in the theory can be expressed
in terms of physical quantities (\ref{cons1}, \ref{cons2}). Only $L_{3,0}$
and $M_{0,0}$ have to be estimated numerically (\ref{l30}, \ref{m00}). 
The final expression is
a function of the mixing angle $\theta$, because the determination
of the free parameters through the masses and decay constants does
not fix the value $\theta$, although there are many reasons (see end of
section 4) to expect it to be around $-21^{\circ}$:

\begin{table}[h]
\begin{center}
\begin{tabular}{|c|c|c|c|}
\hline
$\theta$&
$-20^{\circ}$&
$-21^{\circ}$&
$-22^{\circ}$ \\
\hline
$\Gamma (\eta\rightarrow \pi_0 \pi_0 \pi_0)$&
$129.1\; {\rm eV} $& $124.8\; {\rm eV} $& $119.7\; {\rm eV} $  \\
\hline
$\Gamma (\eta\rightarrow \pi_0 \pi_+ \pi_-)$&
$87.9\; {\rm eV} $ & $84.7\; {\rm eV} $& $80.9\; {\rm eV} $ \\
\hline
r &
$1.47$  & $1.47 $& $1.48$ \\
\hline
\end{tabular}
\end{center}
\end{table}

The predicted rates
have improved but remain too small when compared to the experimental data.
This result seems to point out that the pion-pion final-state scattering
effects are indeed a key element in the process. In the context of
$U_L(3)\otimes U_R(3)\ \chi {\rm PT}$, these are next-to-next-to-leading order 
contributions. When properly evaluated, this should include
the one-loop correction that stems from the leading-order Lagrangian, and a 
myriad of new terms ---including some ${\cal O}(p^6)$ terms--- coupled
to totally unknown parameters. 
One expects, however, that the most important contribution should come from
the pion-pion interaction and that it should not be much different from
the $SU(3)$ result \cite{gl2}. 

\vspace{2em}
\section{ The $\eta' \rightarrow \pi \pi \pi$ decays.}

The calculation goes along the same lines as the one presented in the 
previous section
although
there is obviously no $SU(3)$ prediction to compare with. 
The processes are also isospin violating and will be evaluated to first order
in $\epsilon_0$.

The expressions (\ref{amplitudes}), (\ref{ratio}) 
and the vector-meson dominance
argument also apply to this process, so the 
ratio between the charged and neutral
channels can also be predicted to be around 1.5. This number is compatible
with the experimental data.

\vspace{1em}
\subsection{Leading order }
\vspace{1em}

The leading-order amplitude for the 
$\eta' \rightarrow \pi_0 \pi_+ \pi_-$ decay reads:
\bea
A_{0+-}&=& \frac{\epsilon_0}{f_{\pi}} \ \tth \ \frac{\st+\sqrt{2} \ct}{\st-
\sqrt{2} \ct} \; \Big( (s - \frac{4}{3} M^2_{\pi}) \
(\sqrt{2} \ct+\st) \nonumber \\
&-& \frac{M^2_{\pi}}{3}\ \ctt \ 
(4 \sqrt{2} \st-\ct)
\Big) \; .
\label{pepm}
\eea

Notice that $\theta \rightarrow \theta + \frac{\pi}{2}$ exchanges $\eta$
and $\eta'$ (except for a global sign) {\em only} in the isospin limit,
because the $\pi-\eta$ and $\pi-\eta'$ mixings are not equal, so this
symmetry cannot be used to relate (\ref{eopm}) and (\ref{pepm}). 

The integration over the allowed phase-space region leads to:
\bea
\Gamma (\eta'\rightarrow \pi_0 \pi_0 \pi_0) &=& 457  \; {\rm eV} \ ,\nonumber \\
\Gamma (\eta'\rightarrow \pi_0 \pi_+ \pi_-) &=& 405 \; {\rm eV} \ ,\nonumber \\
r&=& 1.13\; . \nonumber 
\eea
The prediction is of the correct order of magnitude. These results agree
with the estimations in \cite{af}. Notice, in particular, that the present
computation does include the effects of $\eta_0-\eta_8$ mixing and the
gluon anomaly that they point out as a crucial ingredient of their
calculation. 

\vspace{1em}
\subsection{Next-to-leading order }
\vspace{1em}

The analytical expressions for the amplitudes are given in the appendix.
It turns out that the numerical results are not good, 
because the corrections are huge:
\begin{table}[h]
\begin{center}
\begin{tabular}{|c|c|c|}
\hline
$\theta$&
$-20^{\circ}$&
$-22^{\circ}$ \\
\hline
$\Gamma (\eta'\rightarrow \pi_0 \pi_0 \pi_0)$&
$2280\; {\rm eV} $& $2137\; {\rm eV} $ \\
\hline
$\Gamma (\eta'\rightarrow \pi_0 \pi_+ \pi_-)$&
$1642\; {\rm eV} $ & $1536\; {\rm eV}  $ \\
\hline
r &
$1.39$  & $1.39 $ \\
\hline
\end{tabular}
\end{center}
\end{table}

One would {\em a priori} expect close similarities between both $\eta$ and
$\eta' \rightarrow \pi \pi \pi$ processes. The $\eta'$ is however
nearly twice as massive as the $\eta$, so the emerging pions will have
relatively high momenta. This might suggest that the bad
results quoted above are to be blamed on a breakdown of the low-energy
expansion. However, high momenta also allow for a lot of phase space
for pion-pion interactions that produce intermediate resonances
like the $\rho$ or the $\sigma$. This suggests that, in this case,
the effect of resonances might be even more important that in the
$\eta \rightarrow \pi \pi \pi$. This would correspond to one-loop
corrections, but 
one can barely expect a third contribution to cancel the
huge second-order corrections in order to produce a good final result.

\vspace{2em}

\section{ The $\eta' \rightarrow \eta \pi \pi$ decays.}

This transition is also described by the Lagrangian discussed in
section 4 and will not include 
new elements to be taken into account.
As a matter of fact, the calculation is simpler because these 
are not isospin-violating processes so one can set $\epsilon_0=0$ everywhere.

\vspace{1em}
\subsection{Leading order}
\vspace{1em}

The leading-order amplitudes for both the charged and the neutral
channel read:
\bea
A_{\eta' \rightarrow \eta \pi \pi}(s,t,u)&=& \frac{M^2_{\pi}}{6\ f_{\pi}}
\Big( 2 \sqrt 2 \cos( 2 \theta) - \sin (2 \theta)\Big) \ .
\label{b-bo}
\eea

The decay rates that follow from (\ref{b-bo}) are however very small
\cite{fg,oneida,vdv}:
\bea
\Gamma (\eta'\rightarrow \eta \pi_0 \pi_0) &=&  1.0 \; {\rm keV} \ ,\nonumber \\
\Gamma (\eta'\rightarrow \eta \pi_+ \pi_-) &=&  1.9\; {\rm keV} \ ,\nonumber \\
r&=&1.9\; . \nonumber
\eea

The actual values for the rates are however 40 times 
smaller than the experimental ones. A possible
justification for the leading-order prediction being so small was already 
pointed out in \cite{vdv,akhoury}: notice that the leading-order amplitude
(\ref{b-bo}) vanishes in the chiral limit. 

\vspace{1em}
\subsection{Next-to-leading order }
\vspace{1em}

The first non-vanishing
contribution in the chiral limit would come  
from the next-to-leading Lagrangian
that does not break chiral symmetry explicitly: $O_0$ and $O_3$. 
The actual computation shows that  
even when massive quarks are considered, the 
most important contributions to the amplitude do indeed come from 
these operators. Numerically the contribution from $O_3$ is suppressed
by the small value of $M_{3,0}$ (\ref{m30}). 
In the center of the Dalitz plot, where
$s=t=u=\frac{M^2_{\eta'}+M^2_{\eta}+2\ M^2_{\pi}}{3}$,
the contribution from $O_0$ to the total amplitude
is equal to 10 times the leading-order amplitude.
This seem to indicate that the key mechanism associated to this
transition is related to some qualitatively special dynamics that
are not reflected in the leading-order Lagrangian, but show up
in the next order (and, in particular, in the $O_0$ operator). 
As in the previous sections, the analytical expression for
the amplitudes has been relegated to the appendix.

As a consequence of this strong dependence on one particular 
operator, these processes provide an excellent way to fix the
unknown constant $M_{0,0}$. In this work, the fitting was done
in the charged channel (\ref{m00}). 

From a strict perturbative point of view, only linear
next-to-leading corrections should appear in the decay rate. This
implies that the squared absolute value of the amplitude is
introduced in an expanded form: $|A_{LO}|^2+ 2 |A_{LO}||A_{NLO}|$.
In the case that is considered in this section, however, 
this expansion does not make sense,
since the second-order contribution
to the amplitude is by no means a small correction to the leading-order
value! The relevant dynamics appear in the next-to-leading Lagrangian and 
should not be treated as a perturbation. The decay rate
must be computed from the square of the absolute value of the amplitude,
$|A_{LO}+A_{NLO}|^2$,
and not from an expanded form. Otherwise one obtains a 
negative value for the rate! This is not a sign of a poorly convergent
expansion, because third- and  higher-order corrections could still be expected
to be small. The point is that the leading-order
contribution is (nearly) zero, so the next-to-leading contribution is
actually to be
considered as the first term in the expansion.
 
The results are also a function of the mixing angle, but 
the dependence is negligible in the region of interest, 
$-20^{\circ}\leq\theta\leq-22$: 

\bea
\Gamma (\eta' \rightarrow \eta \pi_0 \pi_0) &=& 48.7 \; {\rm eV} \; ,\nonumber \\
\Gamma (\eta' \rightarrow \eta \pi_+ \pi_-) &=& 88.9 \; {\rm eV} \; ,\nonumber \\
r &=& 1.83 \; . \nonumber
\eea

The agreement between the predicted and the experimental value 
for the neutral channel is almost within the experimental error.
The ratio between the two channels also matches the measured value. 
A word of caution is however needed in relation to 
these results: the decay rates turn out to be extremely
sensitive to the value of this constant. Some illustrative figures: 
a deviation of $+0.05 \cdot 10^{-3}$ in $M_{0,0}$ increases
the estimated rates by 30\%. The ratio remains almost unaffected,
as dictated by isospin symmetry. 
 
Some recent work on a pseudoscalar-scalar meson coupling model \cite{amir}
(see also \cite{dt})
indicates that the dominant contribution to 
$\eta' \rightarrow \eta \pi \pi$ decays comes from the exchange of
the intermediate scalar resonance $a_0(980)$. It also predicts 
a less important contribution from the $\sigma$, and no tree-level 
contribution from the $\rho$ (forbidden by G-parity). These features
agree from a qualitative point of view 
with the results in this paper,
since the $a_0$ resonance does indeed contribute to $M_{0,0}$ \cite{drp}. 
The $\rho$ internal exchanges are a delicate issue in the $U(3)$
formalism, because $M_{\eta'}>M_{\rho}$, which means that the $\rho$
particle has not been fully integrated out.
The G-parity constraint protects the $\eta' \rightarrow
\eta \pi \pi$ system against this obstacle.
The only relevant (but smaller) one-loop corrections should 
correspond to $\sigma$ exchanges.


\section{Conclusions}

The $U_L(3) \otimes U_R(3)$ formalism provides a systematic way of
dealing with the $\eta - \eta' -\pi$ interactions. In particular,
it is supposed to give a fairly accurate description of the
$\eta /\eta'$ system.
The good results obtained in \cite{meta3} for the values of the masses and the
mixing angle $\theta$ are also certainly a strong motivation for the study
of these decays. 

The effective theory is built upon the assumption that all
particles other than the nine Goldstone bosons can be integrated out,
so their effects would show up in the effective vertices. This assumption
is exact in the $m_q \rightarrow 0$ {\em and} $N_c \rightarrow \infty$ limit,
because then the bosons are really massless and any massive resonance
will decouple.
As we move away from this ideal situation, quark-mass and 
$1/N_c$ perturbations are introduced. This produces three
sort of corrections. Firstly, new interactions between the Goldstone
bosons themselves appear (consider, for instance, the leading-order
mass terms, that are either of ${\cal O}(m_q)$ or ${\cal O}(1/N_c)$). 
Secondly, the integration of all intermediate \-resonances
becomes less reliable. Light resonances like
$\rho$, $\omega$ and $\sigma$ can indeed produce serious \-problems,
specially when they happen to play an important role in some particular
process. Whenever this occurs, one-loop contributions, where these 
resonances appear in terms of interacting Goldstone bosons,  
ought to be essential.
This seems to be certainly the case for $\eta \rightarrow \pi\pi\pi$
decays and probably for
$\eta' \rightarrow \pi\pi\pi$ decays, too. 
(Notice that the integration of $\rho$ and $\omega$ is more delicate
in the $U(3)$ formalism than it was in the octet theory, 
because $M_{\eta'}$ is around 1 GeV, which is 
{\em higher} than the masses of these particles, so their low modes
have not been integrated out. 
This might imply a change in the estimated
value of some constants like $L_{3,0}$ because $L_3$ feeds precisely on 
isovector-meson contributions \cite{drp}).
In the third place, 
the anomalous diagrams involving $g^2 \ 
G\tilde{G}$ ---the gluonic part of $\eta_0$ (see figure) must be
associated to next-to-leading contributions, because they are 
suppressed in $1/N_c$ with respect to the diagrams involving quark currents.
\begin{figure}[here]
\epsfig{file=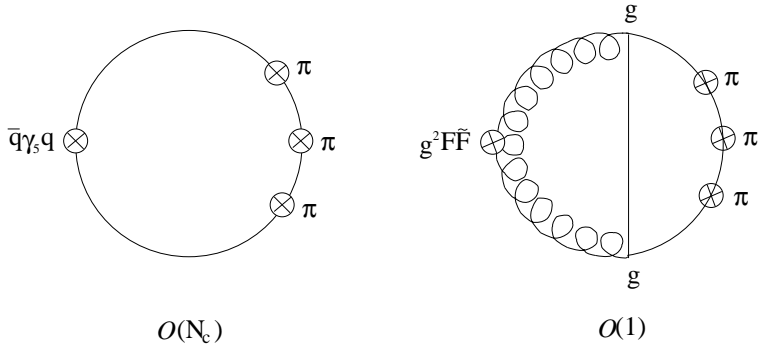,width=100mm}
\caption{Quark-current and glueball contributions in 
$\eta \rightarrow \pi \pi \pi$. A similar diagram can be drawn for
$\eta'$ decays.}
\end{figure}

The $U(3)$ formalism offers a more complete description of the
$\eta$ than the octet theory, because the latter identifies
$\eta_8$ and $\eta$, neglecting mixing effects
that should to be rather important because the mixing angle is not
that small. 
The $U(3)$ calculation
improves the tree-level $SU(3)$ prediction for
the $\eta \rightarrow \pi \pi \pi$ decays. 
The most relevant new feature for the
these decays seems to be the $\eta-\eta'$ mixing 
(the anomaly does play a role there but
it is not a direct contribution to the 4-point function). 
The mixing angle is well-predicted at
leading order and is stable under next-to-leading
corrections ($\theta \approx -20^{\circ}/-22^{\circ}$), 
so significant corrections
due to mixing effects are not expected to appear at higher orders.
Nevertheless, since the next-to-leading 
predicted rates stay well below the experimental
values, there seems to be no way to avoid the importance
of unitary corrections and low-energy resonances
whose effects are not included in the effective vertices precisely
due to their small mass. 
The ${\cal O}(\delta^3)$ effective action would include these
final-state-interaction corrections, 
but tadpole corrections as well as many new terms in the
Lagrangian with unknown coupling constants should also be taken into account.
If the discrepancies with experimental values 
are indeed mainly due to the presence of
intermediate states, tadpoles and ${\cal O}(\delta^3)$ counterterms 
could be safely 
neglected and the unitary corrections ---that can be computed with the
known ingredients--- should be the only contribution that matters. 
This one-loop calculation is however out of the scope of the present 
article and 
will be left for future work. (A first step in this 
direction is the evaluation of
the one-loop contributions to the masses and decay constants \cite{kaiser}). 

The same discussion should also apply to the $\eta'\rightarrow \pi \pi \pi$
decays. Although the leading-order approximation produces good
estimates of the rates, the emerging pions are far from being soft and
this is probably the reason why the expansion apparently blows up when 
corrections are included.  

$\eta' \rightarrow \eta \pi\pi$ transitions are probably 
the most interesting decays in this paper, because 
this decay could not be analyzed in the framework of $SU(3)$.
Momenta are not expected to be too high, so the 
low-energy expansion might actually
work. The results are reasonably good at tree level, but an 
estimate of the one-loop corrections
would also be of great interest in this case, in order to check
the convergence of the expansion, which is certainly one the most
fragile points in the $U(3)$ theory.

\vspace{2em}

\section{Acknowledgments}
\vspace{1em}

The author is thankful to J. I. Latorre and J. Taron for suggesting the
problem, for their critical reading of the manuscript and for many discussions
and comments. She is also grateful to P.~Pascual, who took part
in some of these discussions and shared his wide experience in the field
of particle physics. She also thanks comments and questions from 
A. Fariborz and D.~Black, who also helped with english corrections.
The Physics Department of Syracuse
University is acknowledged for its hospitality.
The project is financially supported by CICYT, contract AEN95-0590,
and by CIRIT, contract GRQ93-1047, and by a grant from the 
{\em Generalitat de Catalunya}.


\begin{appendix}

\vspace{1em} 
\section{Isospin symmetry for $\eta / \eta' \rightarrow \pi\pi\pi$ decays}
\vspace{1em} 

The relation (\ref{amplitudes}) between the charged and the neutral 
channels can be proved by simply considering
isospin symmetry. By definition, for the charged channel,
\bea
A(s,t,u) &=& \langle \pi_0(p_1)\ \pi_+(p_2)\ \pi_-(p_3)\ | 
\ \eta(p_4) \rangle \; ; \nonumber \\
A(t,u,s) &=& \langle \pi_0(p_2)\ \pi_+(p_3)\ \pi_-(p_1)\ | 
\ \eta(p_4) \rangle \; ; \nonumber \\
A(u,s,t) &=& \langle \pi_0(p_3)\ \pi_+(p_1)\ \pi_-(p_2)\ | 
\ \eta(p_4) \rangle \; . \nonumber 
\eea
The permutation of momenta can be viewed as a
permutation in the isospin values instead. For instance
\footnote{The minus signs are due the fact that $(\pi_+)^*= -\pi_-$.}:
\bea
| \pi_0(p_2)\ \pi_+(p_3)\ \pi_-(p_1)\ \rangle &=& 
-| p_1\ p_2\ p_3\ ; \ 11\ 1{-1}\ 10\ \rangle \; ;
\nonumber \\
| \pi_0(p_2)\ \pi_+(p_1)\ \pi_-(p_2)\ \rangle &=& 
-| p_1\ p_2\ p_3\ ; \ 11\ 10\ 1{-1}\ \rangle \; ; \nonumber \\
\vdots \nonumber
\eea
 
These three particle states can be expressed in the Clebsch-Gordan 
basis
$| I^{(2,3)}\ I\ M \rangle  $
where the states are labeled in terms of   
the isospin from particles 2 and 3 $I^{(2,3)}$,  
the total isospin $I$ and the component in the $z$-direction
of the total isospin $I_z$. 

Due to the bosonic nature of pions, only the totally symmetric
states will contribute to the decay amplitude. This
restricts the final state to a superposition of three states
(expressed in the Clebsch-Gordan basis):
\bea
| symmetric\, state \rangle = \frac{6}{\sqrt{10}}|2\ 3\ 0  \rangle + 
\frac{4}{\sqrt{15}} | 2\ 1\ 0 \rangle  + \frac{2}{\sqrt{3}}
| 0\ 1\ 0 \rangle \; .
\label{sym1}
\eea

The neutral-channel analysis is much simpler, since all particles
are $I=1, \, I_z=0$, which can only produce totally
symmetric states:
\bea
| 10 \ 10 \ 10 \rangle &=& \sqrt{\frac{2}{5}} | 2\ 3\ 0 \rangle -
\frac{2}{\sqrt{15}} | 2\ 1\ 0 \rangle -
\frac{2}{\sqrt{3}} | 0\ 1\ 0 \rangle \; .
\label{sym2}
\eea

However, the $\eta \rightarrow \pi\pi\pi$ transition must be induced by 
the isospin-violating piece in the QCD Lagrangian \cite{gl2}: 
\bea
{\cal L}_{{\rm QCD}} = -\frac{1}{2}(m_u-m_d) \ (\bar u u- \bar d d) \; .
\nonumber
\eea
This is a $\Delta I$=1 operator, so the $I$=3 
pieces in (\ref{sym1}) and (\ref{sym2}) 
will not contribute to the decay
amplitude. The remaining terms differ by the value of 
$I^{(2,3)}$, so this number can be used to label the 
two different contributions to the charged amplitude:
\bea
A(s,t,u) &=& \langle \ 10\ 11\ 1{-1}\ | \ 00 \
\rangle \; = \; \frac{1}{\sqrt 3}\ A_0 - \frac{1}{\sqrt 15}\ A_2
\; ; \nonumber \\
A(t,u,s) &=& \langle \ 11\ 1{-1}\ 10\ | \ 00 \
\rangle \; = \; \sqrt{\frac{3}{20}}\ A_2 
\; ; \nonumber \\
A(s,u,t) &=& A(s,t,u) 
\; ; \nonumber \\
A(u,t,s) &=& A(u,s,t) \; =\; A(t,s,u) \; =\; A(t,u,s) 
\; ; \label{amp1}
\eea
and to the neutral amplitude:
\bea
\bar A(s,t,u) &=& \langle \ 10\ 10\ 10\ | \ 00 \
\rangle \; = \; - \frac{1}{\sqrt 3}\ A_0 - \frac{2}{\sqrt 15}\ A_2
\; .
\label{amp2} 
\eea 

From (\ref{amp1}) and (\ref{amp2}), it is straightforward to check that:
\bea
\bar A (s,t,u)&=& A(s,t,u)+A(t,u,s)+A(u,s,t) \; . \nonumber
\eea

Obviously the analysis applies to both $\eta$ and $\eta'$ 
decays.
  
\vspace{1em}

\section{Next-to-leading decay amplitudes}

The analytical expressions of the decay amplitudes are lengthy 
due to the dependences on the mixing angle $\theta$.
As discussed in the main body of this article, every constant 
in the computation except $M_{0,0}$ and $L_{3,0}$ 
can be written in terms of measurable quantities and $\theta$:
\bea
\Delta_M &=&  \frac{3\ M^2_{\eta}+M^2_{\pi}-4\ M^2_K+3\ \sin^2 \theta
\ (M^2_{\eta'}-M^2_{\eta})} {4\ (M^2_K-M^2_{\pi})} \; ; \nonumber \\
\Delta_N &=& \frac{3}{4\ \sqrt 2} \
\frac{M^2_{\eta'}-M^2_{\eta}}{M^2_K-M^2_{\pi}}\ \sin 2\theta \; ; \nonumber \\
\Delta_P &=& \big( \frac{f_K}{f_{\pi}}-1 \big)\; ; \nonumber \\
v_{4,0} &=&  -\frac{2}{3}\ \frac{f_{\eta}+f_{\eta'}}{f_{\pi}}\ 
-\ 2\; ; \nonumber \\
\eea 

The amplitudes will always have the general form:
\bea
A_{NLO}(\theta) &=& A_{LO}(\theta) \ + \Delta_M(\theta) A_M(\theta) 
+ \Delta_N(\theta) A_N(\theta) + \Delta_P A_P(\theta) + 
v_{4,0} A_{4,0}(\theta) \nonumber \\ &+& 
\frac{M_{0,0}}{f^2} A_{0,0}(\theta) + \frac{L_{3,0}}{f^2} A_{3,0}(\theta) 
\; , \nonumber
\nonumber
\eea

The actual expressions for these next-to-leading contributions 
to the charged-channel decay amplitudes are given below. The 
amplitudes for the neutral channels can be inferred from them.

\vspace{1em}
\subsection{$\eta \rightarrow \pi\pi\pi$}
\vspace{1em}

\bea
A_M &=& -\frac{2}{3\cd\ (\sqrt 2 \ct+\st)^2(\ct-\sqrt 2 \st)} 
\times \nonumber \\
&& \Big( \
3 s \cd \ (-2 +7\cd - 7\cq - 2\sqrt 2 \st \ct + 4 \st \ccu) \nonumber \\
&+& M^2_{\pi} \ct \ (35 \ct -104 \ccu + 85 \cc + 5 \sqrt 2 \st + 
12 \sqrt 2 \st\cd \nonumber \\ &-& 
37 \sqrt 2 \st \cq) 
+ \frac{M^4_{\pi}}{M^2_{\eta}-M^2_{\pi}} \ ( -2-40\cd+
106 \cq - 64 \csx \nonumber \\ &-& 13 \sqrt 2 \st\ct  
+ 4 \sqrt 2 \st \ccu + 25 \sqrt 2 \st \cc )\
\Big) 
 \; , \nonumber \\
A_N &=& \frac{1}{3(\sqrt 2 \ct+\st)(\ct-\sqrt 2 \st)} \times \nonumber \\
&& \Big( -3 s \sin 2\theta \ 
(\cos 2\theta +\frac{\sqrt 2}{4} \sin 2 \theta )+
 M^2_{\pi} \sqrt 2 \ (1 +16 \cd - 17 \cq) \nonumber \\
&+&  10 M^2_{\pi} \st \ (3-5 \cd) \
\Big) 
  \; , \nonumber \\
A_P &=&  -\frac{4}{9\cd \st \ (\st+\sqrt 2 \ct)^2(\ct -\sqrt 2 \st) } \
\frac{M^2_{\pi}}{M^2_{\eta}-M^2_{\pi}} \times \nonumber \\ 
&&\Big( \ 3s\ (-2 \sqrt 2 \ct + 
5 \sqrt 2 \ccu - 2 \sqrt 2 \cc - \sqrt 2 \csv \nonumber \\ &-&
8 \cd \st+ 26 \cq \st- 22 \csx \st) \nonumber \\
&+& M^2_{\pi} \ (21 \sqrt 2 \ct  -37 \sqrt 2 \ccu - 13 \sqrt 2 \cc + 29
\sqrt 2 \csv  \nonumber \\ &+& 2 \st +72 \cd \st 
- 210 \cq \st + 152 \csx \st  )
\Big)
\; , \nonumber \\
A_{4,0} &=& -\frac{3}{2}\ A_N
\; , \nonumber \\
A_{0,0} &=& 8\ \st \ (4\sqrt 2 \ct + \st)\ \Big( s^2-3 s s_0 - u t + 
3 M_2^2  \Big)\; , \nonumber \\
A_{3,0} &=& \frac{4}{3 \ct(\ct-\sqrt 2 \st)} \Big( \ \ct \ (s^2 -
15 s s_0+2 t u) \nonumber \\
&-& \ct \sin^2 \theta \ (42  M_2^2
+13 s^2+57s s_0-16 tu) 
- 2 \sqrt 2 \st \ (
3 M_2^2 +s^2 +3 s s_0 -tu) \nonumber \\ 
&+& 2 \sqrt 2 \st \cos^2 \theta \ (15  M_2^2 
+4 s^2+30 s s_0 -7 tu)  
\Big)
\; , \nonumber 
\eea
where $s_0=M^2_{\eta}/3+ M^2_{\pi} $ and $M_2^2=
M^2_{\pi}\ (M^2_{\eta}+M^2_{\pi})$.

\vspace{1em}
\subsection{$\eta' \rightarrow \pi\pi\pi$}
\vspace{1em}

\bea
A_M &=& \frac{2}{9 \ct \st (\ct - \sqrt 2 \st)^2}\times \nonumber \\ 
&&\Big(   9 s\  
    (-3 \ct + 8\ccu - 5 \cc -  \sqrt 2  \st + 2 \sqrt 2  \cd \st - 
       \sqrt 2  \cq \st) \nonumber \\  &+&  
 \frac{4\ M^4_{\pi}}{M^2_{\eta'}-M^2_{\pi}} \ 
(-16 \ct + 56 \ccu - 38 \cc + 2 \sqrt 2  \cd \st - 
      13 \sqrt 2  \cq \st) \nonumber \\  &+& M^2_{\pi} 
    (92 \ct - 267 \ccu + 175 \cc + 28 \sqrt 2  \st - 71 \sqrt 2  \cd \st 
\nonumber \\  &+&  
      62 \sqrt 2  \cq \st)\Big)
\; , \nonumber \\
A_N &=& \frac{1}{3 (\ct-\sqrt 2 \st)} \Big( 
M^2_{\pi}\ (\sqrt 2 \cd-10 \sqrt 2 \sd -28 \st \ct) \nonumber \\ &+&
3 s \st \ (\sqrt 2 \st +2 \ct)
\Big)
\; , \nonumber \\
A_P &=& -\frac{4}{9 \st\ct (\ct-\sqrt 2 \st)^3}\ 
\frac{M^2_{\pi}}{M^2_{\eta'}-M^2_{\pi}}\times \nonumber \\
&&\Big(M^2_{\pi}\ (-16 + 
108 \cd - 246 \cq + 152 \csx - 
        24 \sqrt 2  \ct \st 
\nonumber \\ &+& 74 \sqrt 2  \ccu \st - 29 \sqrt 2  \cc \st) - 
     3 s (-4 + 22 \cd - 40 \cq + 22 \csx \nonumber \\ &-& 2 \sqrt 2  \ct \st  
       + 5 \sqrt 2  \ccu \st - \sqrt 2  \cc \st)
\Big)
\; , \nonumber \\
A_{4,0} &=& - \frac{3}{2}\ A_N\; , \nonumber \\
A_{0,0} &=& \frac{8 (\st+\sqrt 2 \ct)(4 \sqrt 2 \st -\ct)}{\ct -\sqrt 2 \st} \
\Big( s^2-3 s s_0 - u t + 
3 M_2^2 \Big)\; , \nonumber \\
A_{3,0}  &=& \frac{4}{3 \ct(\ct-\sqrt 2\st)} \ \Big(
- \st\ (s^2-15 s s_0 +2 tu) \nonumber \\ &+&
\st \cd \ (42 M_2^2 +13 s^2+57 s s_0 -16 tu)
-2 \sqrt 2 \st \ (3 M_2^2+s^2+3 s s_0 -tu )\nonumber \\ &+&
2 \sqrt 2 \ct \sd \ (15M_2^2+4 s^2+30 s s_0 -7tu )
\Big)
\; , \nonumber 
\eea
where $s_0=M^2_{\eta'}/3+ M^2_{\pi} $ and $M_2^2=
M^2_{\pi}\ (M^2_{\eta'}+M^2_{\pi})$.

\vspace{1em}
\subsection{$\eta' \rightarrow \eta\pi\pi$}
\vspace{1em}

\bea
A_M &=&  -\frac{8}{9 \ct} \ \frac{M^4_{\pi}}{M^2_{\eta}-M^2_{\pi}} \
	(2 \sqrt 2  \ccu + 2 \scu -
       3 \sqrt 2  \ct \sd )
\; , \nonumber \\
A_N &=& \frac{M^2_{\pi}}{3}\ (\sqrt 2  \cos 2\theta  - 2 \sin 2\theta)
\; , \nonumber \\
A_P &=& \frac{2}{9 \ct (\sqrt 2 \ct+\st)} \ 
\frac{ M^2_{\pi}}{M^2_{\eta}-M^2_{\pi}} \
 \Big(4 M^2_{\pi} - 
     2 \cd\ (3 M^2_{\eta} - 6 M^2_{\eta'} + M^2_{\pi})  \nonumber \\ &+&
4 \cq\ (3 M^2_{\eta} - 3 M^2_{\eta'} - 4 M^2_{\pi}) + 
	    \sqrt 2  \ct \st\ (3 M^2_{\eta'} + 7 M^2_{\pi}) \nonumber \\ &+&  
     \sqrt 2 \ccu \st\ (-3 M^2_{\eta} + 3 M^2_{\eta'} - 14 M^2_{\pi}) \Big)
\; , \nonumber \\
A_{4,0} &=& -\frac{3}{2}\ A_N\; , \nonumber \\
A_{0,0} &=& -4\ (2\sqrt 2 \cos 2\theta  - \sin 2\theta)\  
   \Big( M^4_{\pi} + 2 M^2_{\pi}(M^2_{\eta}+M^2_{\eta'})+ 
	M^2_{\eta} M^2_{\eta'} + s^2 - tu + 3 s s_0 \Big)
\; , \nonumber \\
A_{3,0} &=& \frac{1}{3}\ A_{0,0}\; , \nonumber 
\eea
where $s_0=(M^2_{\eta'}+M^2_{\eta}+2 M^2_{\pi})/3 $.

\end{appendix}

\end{document}